\documentclass[dvips]{article}
\usepackage{graphicx}
\usepackage{amssymb}
\usepackage{graphicx}
\usepackage{dcolumn}
\usepackage{amsmath}
\usepackage{lscape}
\usepackage{longtable}
\usepackage{subfigure}
\usepackage{color}

\begin{document}
\newcommand{\bd}{\begin{document}}
\newcommand{\ed}{\end{document}}
\newcommand{\bc}{\begin{center}}
\newcommand{\ec}{\end{center}}
\newcommand{\bfr}{\begin{flushright}}
\newcommand{\efr}{\end{flushright}}
\newcommand{\lt}{\left}
\newcommand{\rt}{\right}
\newcommand{\vs}{\vspace}
\newcommand{\hs}{\hspace}
\newcommand{\beq}{\begin{equation}}
\newcommand{\eeq}{\end{equation}}
\newcommand{\lb}{\linebreak}
\newcommand{\pb}{\pagebreak}
\newcommand{\mb}{\makebox}
\newcommand{\fb}{\framebox}
\newcommand{\mc}{\multicolumn}
\newcommand{\ben}{\begin{enumerate}}
\newcommand{\een}{\end{enumerate}}
\newcommand{\bit}{\begin{itemize}}
\newcommand{\eit}{\end{itemize}}
\newcommand{\ol}{\overline}
\newcommand{\un}{\underline}
\newcommand{\lefq}{\lefteqn}
\newcommand{\ba}{\begin{array}}
\newcommand{\ea}{\end{array}}
\newcommand{\beqa}{\begin{eqnarray}}
\newcommand{\eeqa}{\end{eqnarray}}
\newcommand{\beqas}{\begin{eqnarray*}}
\newcommand{\eeqas}{\end{eqnarray*}}
\newcommand{\bfg}{\begin{figure}}
\newcommand{\efg}{\end{figure}}
\newcommand{\bds}{\begin{displaymath}}
\newcommand{\eds}{\end{displaymath}}
\newcommand{\btb}{\begin{tabbing}}
\newcommand{\etb}{\end{tabbing}}
\bc {
\textbf{\large \textbf{ Spin $1/2$ Particle Dynamics  Near Reissner-Nordstr\"{o}m Black Holes: Extremal and Non-Extremal Cases}  } } \\
\vspace*{1.0cm}
\large \"{O}zlem Ye\c{s}ilta\c{s}
\end{center}
\noindent \\
$\ddagger$ Gazi University, Faculty of Science, Physics Department, 06500, Teknikokullar, Ankara, TURKEY, e-mail: yesiltas@gazi.edu.tr \\ \\

\vspace*{.5cm}
\begin{abstract}
\noindent
In this work, the fermion behaviour near the Reissner-Nordstrom black holes  in quantum mechanics is studied and exact solutions of the Dirac equation for Reissner-Nordstr\"{o}m Black Holes are obtained in terms of special functions. We have obtained different vector potentials which satisfy the solutions.
\end{abstract}
\noindent {\bf keyword:}   Dirac equation, curved space-time, black holes \\


\section{Introduction}
Black holes have their roots in 1915, shortly after Albert Einstein published his theory of General Relativity that is first modern solution of general relativity that would characterize a black hole was found by Karl Schwarzschild in 1916 \cite{1}. In later times, some
observations have confirmed black holes as actual astrophysical objects. After the work by Hawking in  1972 \cite{2}, entitled with Black Holes in
General Relativity, an avalanche of interest in black holes has begun. Meeting of two strong theories as quantum mechanics and general relativity
results in conflicts such as the structure of elementary particles and the particles have to be  represented as an extended field model  for
compatibility with the stress-energy tensor of Einstein field equations. By the way, Recently, the demand for the relationships between black holes
and fundamental particles has been constantly increasing and new information has been obtained. For instance, the Compton wavelength is equal to the
wavelength of a photon whose energy is the same as the mass of that particle.  If we consider black holes as particles, Compton wavelength $\lambda$ of
the black hole as determined by the distant observer using the asymptotic mass $M$; $\lambda=\frac{\hbar}{c M_{Pl}/2}=2R_{S}$. The diameter of the black
hole is $2R_S$, this length becomes identical to the Compton wavelength. Because for all types of black holes( neutral, charged or rotating), the
horizon mass is always equal to twice the irreducible mass observed at infinity according to the the horizon mass theorem \cite{3}. If the horizon mass
of a black hole is taken to be the Planck mass $M_{Pl}$, then a distant observer will find the black hole to have an asymptotic mass $M_{Pl}/2$ and
therefore it is resulting that the  Schwarzschild radius is  equal to the Planck length. In case of Kerr--Newman black hole solutions, the radius of the
Kerr singular ring corresponds to the reduced Compton wave length of the particle \cite{4}. Besides the compatible   research efforts, there is another
field of interest which is known as bound state solutions for the particle states near black holes. Moreover, the electron levels can be located
not only beyond the black hole but also   under the Cauchy horizon \cite{5}. If the charge of the black hole is less than its mass, then it's geometry  has an outer   and   inner horizons. The Reissner--Nordstr\"{o}m (RN)  metric emerged hypothetically as the static solution of Einstein's field equations, including Maxwell's equations where the time component of the vector potential doesn't vanish. It corresponds to a charged but non-rotating spherical gravitational field of mass. Among the different  studies; the level shift in RN geometry is studied and gravitational shift of the $1S_{1/2}$ state was found as $10^{-36} Hz$ \cite{6}, non-existence time-periodic solutions of the Dirac equation \cite{7}, the strong cosmic censorship for the Dirac field in the higher dimensional RN-de Sitter black hole \cite{8}, stationary solutions of the Dirac equation in the RN gravitational field  \cite{9} and other interesting studies \cite{10}, \cite{11}, \cite{12}, \cite{bez}.  The object of the present paper is to propose   different solutions of the equation Dirac equation in the metric of the charged RN black hole.

\section{Dirac Equation and Spherical Geometry}
A spherically symmetric metric with a radial dependent function $\lambda(r) $ has a form of \cite{c}
\begin{equation}\label{z1}
  ds^{2}=-\lambda(r)dt^{2}+\frac{1}{\lambda(r)}dr^{2}+r^{2}d\theta^{2}+r^{2}\sin^{2}\theta d\phi^{2}.
\end{equation}
Here, the function $\lambda(r)=1-\frac{2M}{r}+\frac{Q^{2}}{r^{2}}$ and $c=G=\hbar=1$, $M$ is the mass of the black hole, $Q$ is the charge, (\ref{z1})
is known as RN \cite{9}. In case of $|Q| < M$, $\lambda$ has two roots,   $r=r_{\pm}=M \pm \sqrt{M^{2}-Q^{2}}$. For $|Q| > M$, the
metric  describes
a bare singularity without an event horizon and $|Q| = M$ corresponds to an extremal black hole.  The Dirac equation written in a pseudo-Riemannian
space is written as
\begin{equation}\label{z2}
  i\gamma^{\mu}(\partial_{\mu}+iqA_{\mu}+\Gamma_{\mu}-m)\psi=0
\end{equation}
where $A_{\mu}$ is the electromagnetic four potential, $m$ is the electron mass, $\Gamma_{\mu}$ is the spin connection and $e^{\mu}_{(a)}$ are the tetrads. The spin connection
can be given by
\begin{equation}\label{z3}
  \Gamma_{\mu}=\frac{1}{4}\gamma^{(a)}\gamma^{(b)}e_{(a)}^{\nu} e_{(b)\nu;\mu}
\end{equation}
and $\gamma^{\mu}=e_{(a)}^{\mu}\gamma^{(a)}$. If we use the tetrad which is given below
\begin{equation}\label{z4}
  e^{(a)}_{\mu}=diag\left(\sqrt{\lambda(r)}, \frac{1}{\sqrt{\lambda(r)}}, r, r\sin\theta\right),
\end{equation}
one can give the Dirac equation which reduces into
\begin{equation}\label{z5}
 \left( \gamma^{(0)}\frac{1}{\sqrt{\lambda(r)}}\left(\partial_{t}+ iqA(r)\right)+\gamma^{(1)
 }\sqrt{\lambda(r)}\left(\partial_{r}+\frac{1}{r}+\frac{\lambda'(r)}{4\lambda(r)}\right)+\gamma^{(2)}\frac{1}{r}\left(\partial_{\theta}+\frac{\cot\theta}{2}\right)+\gamma^{(3)}\frac{1}{r
 \sin\theta}\partial_{\phi}-m\right)\psi=0.
\end{equation}
where the gauge potential is $A_{\mu}=(e \varphi(r),0,0,0) $, $q=e$ is the particle charge (for an electron). If we follow \cite{9} and \cite{b}, we take
\begin{equation}\label{z6}
  \psi(t,r,\theta, \phi)=\frac{e^{-iEt}\bar{\psi}}{\lambda(r)^{1/4}r \sqrt{\sin\theta}}
\end{equation}
and defining the first order, linear differential operator $\textbf{K}$ as
\begin{equation}\label{z7}
  \textbf{K}=\left(\gamma^{2}\partial_{\theta}+\frac{\gamma^{3}}{\sin\theta}\partial_{\phi}\right)\gamma^{(0)}\gamma^{(1)}
\end{equation}
and $\textbf{K}\bar{\psi}=k\bar{\psi}$, $k=0,\pm 1, \pm 2,...$ Taking $\bar{\psi}$  as separated into the radial and angular parts,
\begin{equation}\label{z8}
 \bar{\psi}=Y(\theta, \phi)\left(
                             \begin{array}{c}
                               \phi_{1}(r) \\
                               i\phi_{2}(r) \\
                             \end{array}
                           \right),
\end{equation}
as a result,  one can obtain a radial system as
\begin{equation}\label{z9}
\left(  \sqrt{\lambda(r)}\frac{d}{dr}-\frac{k}{r}\right)\phi_1 (r)+\left(\frac{E+e\varphi(r)}{\sqrt{\lambda(r)}}+m \right)\phi_2(r)=0,
\end{equation}
\begin{equation}\label{z10}
  \left(  \sqrt{\lambda(r)}\frac{d}{dr}+\frac{k}{r}\right)\phi_2(r)-\left(\frac{E+e\varphi(r)}{\sqrt{\lambda(r)}}-m \right)\phi_1(r)=0,
\end{equation}
where we note that the normalization of the radial part is equal to $1$. The system of (\ref{z9}) and (\ref{z10}) lead to
\begin{eqnarray}\nonumber
  -\lambda(r)\frac{d^{2}\phi_1(r)}{dr^{2}}+\left(\lambda(r)\frac{U'(r)}{U(r)}-\frac{\lambda'(r)}{2}\right)\frac{d\phi_1}{dr}+ \left(\frac{k^{2}}{r^{2}}-\frac{k
  \sqrt{\lambda(r)}}{r^{2}}+2mU(r)-U(r)^{2}-\frac{k\sqrt{\lambda(r)}U'(r)}{rU(r)}\right)\phi_1(r) &=& 0 \\ \nonumber
  -\lambda(r)\frac{d^{2}\phi_2(r)}{dr^{2}}+\left(\lambda(r)\frac{U'(r)}{U(r)-2m}-\frac{\lambda'(r)}{2}\right)\frac{d\phi_2}{dr}+\left(\frac{k^{2}}{r^{2}}+\frac{k
  \sqrt{\lambda(r)}}{r^{2}}+2mU(r)-U(r)^{2}+\frac{k\sqrt{\lambda(r)}U'(r)}{r(U(r)-2m)}\right)\phi_2(r) &=& 0, \nonumber
\end{eqnarray}
where
\begin{equation}\label{Ur}
  U(r)=\frac{E+e\varphi(r)}{\sqrt{\lambda(r)}}+m.
\end{equation}
 Let us use the transformation below,
\begin{eqnarray}
  \phi_i(r) &=& e^{\int^{r} \mu(z) dz} \bar{\phi_i}(r),~~~~i=1,2 \\
  \mu(r) &=& \frac{2\lambda(r)U'(r)+U(r)\lambda'(r)}{4U(r)\lambda(r)}.
\end{eqnarray}
Under this map, we get
\begin{equation}\label{z11}
\begin{split}
  -\lambda(r)\bar{\phi''_1}(r)-\lambda'(r)\bar{\phi'_1}(r)+(\frac{k^{2}}{r^{2}}+2mU(r)-U(r)^{2}-
  \frac{k\sqrt{\lambda(r)}}{r^{2}}-\frac{k\sqrt{\lambda(r)}U'(r)}{rU(r)}+
  \frac{3\lambda(r) U'(r)^{2}}{4U(r)^{2}}-\frac{U'(r)\lambda'(r)}{4U(r)}+\\ \frac{\lambda'(r)^{2}}{16\lambda(r)}-\frac{\lambda
  U''(r)}{2U(r)}-\frac{\lambda''(r)}{4})\bar{\phi_1}(r)=0,
  \end{split}
  \end{equation}
  \begin{equation}\label{z111}
  \begin{split}
      -\lambda(r)\bar{\phi''_2}(r)-\lambda'(r)\bar{\phi'_2}(r)+(\frac{k^{2}}{r^{2}}+2mU(r)-U(r)^{2}-
  \frac{k\sqrt{\lambda(r)}}{r^{2}}+\frac{k\sqrt{\lambda(r)}U'(r)}{r(U(r)-2m)}+\frac{3\lambda U'^{2}}{4(U(r)-2m)^{2}}-\frac{U'\lambda'}{4(U(r)-2m)} \\
  -\frac{\lambda(r)U''(r)}{2(U(r)-2m)}+ \frac{\lambda'(r)^{2}}{16\lambda(r)}-\frac{\lambda''(r)}{4})\bar{\phi_2}(r)=0.
  \end{split}
  \end{equation}
It can be noted that (\ref{z111}) can be obtained from (\ref{z11}) using the replace $m\rightarrow -m$ and that's why we only include the solutions of the system (\ref{z11}).

%%%%%%%%%%%%%%%%%%%%%%%%%%%%%%%%%%%%%%%%%%%%%%%%%%%%%%%%%%%%%%%%%%%%%%%%%%%%%%%%%%%%%%%%%%%%%%%%%%%%%%%%%%%%%%%%%%%%%%%%%%%%%%%%%%%%%%%%%%%%%%%%%%%%%%%%
%%%%%%%%%%%%%%%%%%%%%%%%%%%%%%%%%%%%%%%%%%%%%%%%%%%%%%%%%%%%%%%%%%%%%%%%%%%%%%%%%%%%%%%%%%%%%%%%%%%%%%%%%%%%%%%%%%%%%%%%%%%%%%%%%%%%%%%%%%%%%%%%%%%%%%%%%%%%
\section{Extremal and Non-Extremal Cases}
In the form of a one parameter family obeyed by the extreme RN black holes we have $M=Q$ which means that the black hole is   like a black body with zero temperature. Because there is no surface gravity and the Hawking temperature is equal to zero. $M^{2}<Q^{2}$ stands for the naked singularity where we have  two imaginary roots  at $r = 0$ and no solution to the Einstein equations. A regular black hole with two horizons correspond to $M^{2} >Q^{2}$ where we have two real roots, $r_{+}>r_{-}>0$ and one can construct the Penrose diagrams for the regions at $r>r_{+}$, $r_{+}>r>r_{-}$ and $r_{-}>r>0$. In this study we mainly investigate the Dirac equation solutions for only extreme and non-extremal cases for obtaining solutions in terms of special functions.
\subsection{non-extremal black hole case: $|Q|<M$}
Once we use an approach for a non-extremal black hole, we have $Q\neq 0$, $\lambda \rightarrow \frac{Q^{2}}{r^{2}}$, $k=0$ and $\varphi(r)=\frac{c}{r}$, our system (\ref{z11}) has an   asymptotic solution when $r\rightarrow 0$ which is
\begin{equation}\label{z12}
  \bar{\phi}_{1}(r)= C_1  \sqrt{\frac{2}{(ce+mQ)r}}K\left(2,\sqrt{\frac{2Er}{ce+mQ}}\right),
\end{equation}
where $K(a,b)$ is the modified Bessel function of the first kind. For the case $k\neq 0$, we get
\begin{equation}\label{z13}
  -\chi''(r)+\left(\frac{k^{2}}{Q^{2}}+\frac{3}{4r^{2}}-\frac{kQ}{r}+\frac{(2cemQ-c^{2}e^{2})r^{2}}{Q^{4}}\right)\chi(r)=0,
\end{equation}
where $\chi(r)=\frac{\bar{\phi}_{1}(r)}{r}$. For $c=\frac{2mQ}{e}$, we obtain
\begin{equation}\label{z14}
  \chi(r)= C_1~ Whittaker M\left(\frac{1}{2}, 1, \frac{2kr}{Q}\right)+ C_2~ Whittaker W\left(\frac{1}{2}, 1, \frac{2kr}{Q}\right).
\end{equation}
On the other hand, we may seek for different solutions of  (\ref{z11}). For the values of  $\lambda=\frac{Q^{2}}{r^{2}}$, we get
\begin{equation}\label{z15}
\begin{split}
  -\phi''_{1}(r)+(\frac{2(E+e\varphi(r))+\frac{mQ}{r}+er\varphi'(r)}{mQ+(E+e\varphi(r))r})\phi'_{1}(r)+
  (\frac{k^{2}}{Q^{2}}-\frac{k}{Qr}+\frac{2mr^{2}(m+\frac{r(E+e\varphi(r))}{Q})}{Q^{2}}-\frac{r^{2}(m+\frac{r(E+e\varphi(r))}{Q})^{2}}{Q^{2}}-\\
  \frac{k(E+e\varphi(r)+er\varphi'(r))}{Q^{2}(m+\frac{r(E+e\varphi(r))}{Q})})\phi_{1}(r)=0.
\end{split}
\end{equation}
Now, we focus on (\ref{z15}) through the below cases.
\\
\emph{\textbf{Case I: The solutions in terms of Airy functions: energy dependent potential parameters} }

In order to terminate the coefficient of the first order differential term in (\ref{z15}), we may write
\begin{equation}\label{z16}
  \frac{2(E+e\varphi(r))+\frac{mQ}{r}+er\varphi'(r)}{mQ+(E+e\varphi(r))r}=0,
\end{equation}
and we find $\varphi(r)$ as
\begin{equation}\label{z17}
  \varphi(r)=\frac{C_{1}}{r^{2}}-\frac{mQ}{e}\frac{1}{r}-\frac{E}{e}
\end{equation}
where $C_1$ is the integration constant. Then,(\ref{z15}) becomes
\begin{equation}\label{z18}
  -\phi''_{1}(r)+\left(\frac{k^{2}}{Q^{2}}-\frac{e^{2}C^{2}_{1}}{Q^{4}}+\frac{2emC_1}{Q^{3}}r\right)\phi_{1}(r)=0
\end{equation}
If we take $C_1=E$, $y=\alpha r$, $\epsilon^{2}=\frac{k^{2}Q^{2}-e^{2}E^{2}}{Q^{4}\alpha^{2}}$, $\alpha=\frac{(2emE)^{1/3}}{Q}$ and $z=y+\epsilon^{2}$,
then,
\begin{equation}\label{z19}
  -\phi''_{1}(z)+z\phi_{1}(z)=0
\end{equation}
which is  Airy differential equation and solutions are given by
\begin{equation}\label{z20}
  \phi_1(z)=a Ai(z)+b Bi(z).
\end{equation}
Here, $Bi(z)$ diverges for the larger values of $z$, then we use $\phi_{1}(z)=a Ai(z)$. The boundary conditions imply $\phi_{1}|_{_{z=0}}=0$, thus,
$z=0=y+\epsilon^{2}$ and
\begin{table}[ht]
\caption{Zeros of Airy Function} % title of Table
\centering % used for centering table
\begin{tabular}{c c c c} % centered columns (4 columns)
\hline\hline %inserts double horizontal lines
% Case & Method\#1 & Method\#2 & Method\#3 \\ [0.5ex] % inserts table
%heading
\hline % inserts single horizontal line
1 & -2.33810 \\ % inserting body of the table
2 & -4.08794 \\
3 & -5.52055 \\
4 & -6.78670 \\
5 & -7.94413 \\
6&  -9.02265 \\
7 & -10.04017\\ [1ex] % [1ex] adds vertical space
\hline %inserts single line
\end{tabular}
\label{table:nonlin} % is used to refer this table in the text
\end{table}
boundary condition is satisfied if $\epsilon^{2}_{1}=2.33810$ that implies
\begin{equation}\label{z21}
  E_{1}=\pm \frac{kQ}{e\sqrt{1+\frac{9.3524m^{2}}{Q^{2}}}}.
\end{equation}
Hence, other values for the energy can be easily calculated using the table above.
\\
\emph{\textbf{Case II: The solutions in terms of hypergeometric functions}}
In this case we can take the coefficient of the first order derivative in (\ref{z15}) as different from zero and give,
\begin{equation}\label{z16}
  \frac{2(E+e\varphi(r))+\frac{mQ}{r}+er\varphi'(r)}{mQ+(E+e\varphi(r))r}=\frac{1}{r},
\end{equation}
and we may arrive at  $\varphi(r)$ as
\begin{equation}\label{z17}
  \varphi(r)= -\frac{E}{e}+\frac{C_1}{r}
\end{equation}
where $C_1$ is the integration constant. Now consider the transformation $\phi_1(r)=y(r)\exp \mu(r) \phi_1(r)$, then we get
\begin{equation}\label{z18}
  -\phi''_{1}(r)+\left(\frac{1}{r}-\frac{2y'(r)}{y(r)}-2\mu'(r)\right)\phi'_1(r)+(\frac{k^{2}}{Q^{2}}-\frac{k}{Qr}+\frac{m^{2}Q^{2}-e^{2}C^{2}_{1}}{Q^{4}}r^{2}+
  \frac{y'(r)}{ry(r)}+\frac{\mu'(r)}{r}-\frac{2y'(r)\mu'(r)}{y(r)}-\mu'^{2}(r)-\frac{y''(r)}{y(r)}-\mu''(r))\phi_1(r)=0
\end{equation}
Using the suggestion for the coefficient of the first order derivative
\begin{equation}\label{z18}
  \frac{1}{r}-\frac{2y'}{y}-2\mu'=-(b_1+\frac{b_2}{r})
\end{equation}
which gives
\begin{equation}\label{z19}
  y(r)= r^{\frac{1-b_2}{2}}e^{-(\frac{b_1r}{2}+\mu(r))},
\end{equation}
then we get
\begin{equation}\label{z20}
  -r\phi''_{1}(r)+\left(b_2+b_1r\right)\phi'_{1}(r)+\left(-\frac{b_1b_2}{2}-\frac{k}{Q}+\frac{(1-b_2)(3+b_2)}{4r}+(-\frac{b^{2}_{1}}{4}+\frac{k^{2}}{Q^{2}})r+
  (\frac{m^{2}}{Q^{2}}-\frac{e^{2}C^{2}_{1}}{Q^{4}})r^{3}\right)\phi_{1}(r)=0.
\end{equation}
In order to transform (\ref{z20}) into the  confluent hypergeometric differential equation, we choose the parameters as given below
\begin{equation}\label{z21}
  b_1=1,~~b_2=-3,~~k=\pm \frac{Q}{2},~~C_1=\pm \frac{Q^{2}m}{e}
\end{equation}
that leads to get
\begin{equation}\label{z21}
  \phi''_1(r)+(3-r)\phi'_1(r)+(\frac{b_1 b_2}{2}+\frac{k}{Q})\phi_1(r)=0
\end{equation}
where $\frac{b_1 b_2}{2}+\frac{k}{Q}=-1$ or $\frac{b_1 b_2}{2}+\frac{k}{Q}=-2$. The solutions are given as
\begin{equation}\label{z22}
  \phi_1(r) \sim r^{2} \exp(-\frac{r}{2}) _{1}F_{1}(\frac{b_1 b_2}{2}+\frac{k}{Q},3,r).
\end{equation}
\emph{\textbf{Case 3: The solutions in terms of Whittaker functions}}
In this section we will use the exact definition for the $\lambda(r)$ which is
\begin{equation}\label{z23}
  \lambda(r)= \frac{Q^{2}}{r^{2}}.
\end{equation}
Using (\ref{z23}) and the transformation below,
\begin{equation}\label{z24}
  \phi_1(r)=r^{a}e^{C_1 \int^{r} \mu(r')dr'}\chi_1(r)
\end{equation}
$a,C_2$ are real constants and we obtain
\begin{equation}\label{z25}
\begin{split}
  r\chi_1''(r)+(-1+2a+2C_1 r \mu(r)-\frac{rU'(r)}{U(r)})\chi'_{1}(r)+
   (k/Q+\frac{a(a-2)}{r}-\frac{k^{2}r}{Q^{2}}+\\
   r^{3}(U(r)^{2}-2mU(r))+C_1 \mu(r)(2a-1)+C^{2}_{1}r\mu(r)^{2}-\frac{aU'}{U}+\frac{krU'}{QU}-\frac{C_1 r\mu(r) U'}{U}+C_1 r \mu'(r))\chi_1(r)=0.
  \end{split}
\end{equation}
If we choose
\begin{eqnarray}
  \mu(r) &=& -\frac{1}{2r},~~U(r)=\frac{C_1}{r^{2}},~~a=0
\end{eqnarray}
(\ref{z25}) turns into
\begin{equation}\label{z26}
  r^{2}\chi_{1}''(r)+\left(\frac{1}{4}+\frac{C^{2}_{1}}{Q^{2}}-\frac{k}{Q}r-\left(\frac{k^{2}}{Q^{2}}+\frac{2C_1
  m}{Q^{2}}\right)r^{2}\right)\chi_{1}(r)=0.
\end{equation}
The solutions of (\ref{z26}) are given by
\begin{equation}\label{z27}
  \chi_1(r)=A M_{\alpha,\beta}\left(\frac{2\sqrt{k^{2}+2C_1 m}r}{Q}\right)+ B W_{\alpha,\beta}\left(\frac{2\sqrt{k^{2}+2C_1 m}r}{Q}\right)
\end{equation}
where $\alpha=-\frac{k}{2\sqrt{k^{2}+2C_1 m}},~~\beta=\frac{iC_1}{Q}$, $M_{\alpha,\beta}(z), W_{\alpha, \beta}(z)$ are the Whittaker functions
\cite{whit}. $\nu$ is the quantum number, then,
\begin{equation}\label{z28}
  -\frac{k}{\sqrt{k^{2}+2C_1 m}}=\nu, ~~\nu=1,2,...
\end{equation}
and if we choose the constant $C_1$ as an energy parameter of the system  $C_1=E^{2}$
\begin{equation}\label{z29}
  E_{\nu}=\pm k \sqrt{\frac{1-4\nu^{2}}{8m\nu^{2}}}.
\end{equation}
We also remind that the vector potential component which is energy dependent
 becomes
\begin{equation}\label{z30}
  \varphi(r)= -\frac{E}{e}+\frac{C_1 Q}{er^{3}}-\frac{mQ}{er}.
\end{equation}
\subsection{Extremal black hole case: $|Q|=M$}
\emph{\textbf{Case 1:}} \\
We can consider that case $M=|Q|$ as the extremal black hole.  We take the function $\lambda(r)$ as
\begin{equation}\label{z31}
  \lambda(r)=(1-\frac{M}{r})^{2}.
\end{equation}
We continue with (\ref{z11}) multiplied by $\frac{r^{2}}{M-r}$, then, it becomes
\begin{equation}\label{z32}
  -(M-r)\phi'_{1}(r)+\left(\frac{M}{r}+\frac{(M-r)U'(r)}{U(r)}\right)\phi'_{1}(r)+
  \left(\frac{k}{r}+\frac{k^{2}+r^{2}(2m-U(r))U(r)}{M-r}+\frac{kU'(r)}{U(r)}\right)\phi_1(r)=0.
\end{equation}
Substituting  $U(r)=\frac{C_1}{r}$ in (\ref{z32}) gives
\begin{equation}\label{z33}
  -\phi_1''(r)+\left(-2C_1m+\frac{C^{2}_{1}-k^{2}-2C_1mM}{r-M}\right)\phi_1(r)=0
\end{equation}
and $\phi_1(r)$ are found to be as
\begin{equation}\label{z34}
\begin{split}
  \phi_1(r)=N J\left(-2\sqrt{-C^{2}_{1}+k^{2}+2C_1mM},~~2\sqrt{2C_1 m}(M^{2}-2Mr+r^{2})^{1/4}\right)\Gamma\left(1-2\sqrt{-C^{2}_{1}+k^{2}+2C_1mM}\right)
\end{split}
\end{equation}
where $J_{n}(2\sqrt{2C_1 m}(M^{2}-2Mr+r^{2})^{1/4})$ are the Bessel polynomials. If we take the parameter $C_1$ as energy $C_1=E$, then,
\begin{equation}\label{z35}
  -2\sqrt{-C^{2}_{1}+k^{2}+2C_1mM}=n=-2\sqrt{-E^{2}+k^{2}+2 E mM},
\end{equation}
and the energy is obtained as
\begin{equation}\label{z36}
  E_{n}=\frac{1}{2}(2mM\pm \sqrt{4k^{2}+4m^{2}M^{2}-n^{2}}).
\end{equation}
In this case, vector potential component can be given by
\begin{equation}\label{vp}
  \varphi(r)= -\frac{m+E}{e}-\frac{C_1 M}{er^{2}}+\frac{C_1+mM}{er}.
\end{equation}
\emph{\textbf{Case 2: Solutions in terms of Heun polynomials }}\\
We substitute $r=\frac{M}{z}$ in (\ref{z11}), then,
\begin{equation}\label{z37}
\begin{split}
 -\frac{z^{4}(z-1)^{2}}{M^{2}}\phi''_{1}(z)+\left(-\frac{(z-1)z^{3}(3z-2)}{M^{2}}+\frac{(z-1)^{2}z^{4}U'(z)}{M^{2}U(z)}\right)\phi'_{1}(z)+\\
(\frac{kz^{2}(-1+k+z)}{M^{2}}+2mU(z)-U(z)^{2}-\frac{k(z-1)z^{3}U'(z)}{M^{2}U(z)})\phi_{1}(z)=0.
\end{split}
\end{equation}
After simplification of (\ref{z37}), we use $\phi_1(z)=e^{\int \mu(z)dz}Y(z)$ and get
\begin{equation}\label{z38}
\begin{split}
  Y''(z)+
  \left(\frac{1}{z-1}+\frac{2}{z}+2\mu(z)-\frac{U'(z)}{U(z)}\right)Y'(z)+(\frac{k-k^{2}-kz}{(z-1)^{2}z^{2}}-\frac{2mM^{2}U(z)-M^{2}U(z)^{2}}{(z-1)^{2}z^{4}}+\\
  \frac{\mu}{z-1}+\frac{2\mu}{z}+\mu(z)^{2}+\frac{kU'(z)}{z(z-1)U(z)}-\frac{\mu(z)U'(z)}{U(z)}+\mu'(z))Y(z)=0.
\end{split}
\end{equation}
In order to transform (\ref{z38}) into a Heun equation, we can use the value of $\mu(z)$ given below
\begin{equation}\label{z39}
  \mu(z)= \frac{\epsilon}{z-a}+\frac{1}{2}\frac{2-\gamma+z(\gamma-3+\delta)}{z(z-1)},
\end{equation}
where $ \epsilon, \delta, \gamma$ and $a$ are real parameters, then we obtain
\begin{equation}\label{z40}
\begin{split}
  Y''(z)+(\frac{\delta}{z-1}+\frac{\gamma}{z}+\frac{\epsilon}{z-a})Y'(z)+
  (-\frac{2mM^{2}U(z)}{z^{4}(z-1)^{2}}+\frac{M^{2}U(z)^{2}}{z^{4}(z-1)^{2}}+\frac{(-2+2k+3z)U'(z)}{2z(z-1)U(z)}-\frac{3U'(z)^{2}}{4U(z)^{2}}+\frac{U''(z)}{2U(z)}
  +\\ \frac{1}{4(a-z)^{2}(z-1)^{2}z^{2}}(4a^{2}k(1-k))+a^{2}\gamma(\gamma-2)+z^{4}(\gamma-3+\delta+\epsilon)(1+\gamma+\delta+\epsilon)+\\
  z(-8ak-4a^{2}k+8ak^{2}-2a^{2}(-2+\gamma(-2+\gamma+\delta))-2a\gamma(-2+\gamma+\epsilon))+\epsilon(2\gamma+\delta)+ \\
  z^{3}(4-4k-2(\gamma+\epsilon)(-2+\gamma+\delta+\epsilon)-2a(-3+\gamma^{2}+\delta(-2+\delta+\epsilon)+\gamma(-2+2\delta+\epsilon))))Y(z)=0.
  \end{split}
\end{equation}
Considering the Heun differential equation given in the literare \cite{heun1}, \cite{heun2}:
\begin{equation}\label{z41}
  w''(z)+(\frac{\gamma}{z}+\frac{\delta}{z-1}+\frac{\epsilon}{z-a})w'(z)+\frac{\alpha \beta z-q}{z(z-1)(z-a)}w(z)=0.
\end{equation}
Our task is to adapt (\ref{z40}) into (\ref{z41}) using an appropriate $U(z)$ trial function and parameters. Let us use a constant $U(z)$ as
\begin{equation}\label{z42}
  U(z)=2m
\end{equation}
and this choice makes the vanishing terms including the black hole mass $M$ in (\ref{z40}). From $U(z)=2m$ in (\ref{Ur}), we obtain
\begin{equation}\label{fii}
  \varphi(z)=-\frac{E-m+mz}{e}.
\end{equation}
Then, we can match the coefficient of derivative-free term $Y(z)$ in (\ref{z40}) using the relation below:
\begin{equation}\label{z43}
\begin{split}
 (\frac{(-2+2k+3z)U'(z)}{2z(z-1)U(z)}-\frac{3U'(z)^{2}}{4U(z)^{2}}+\frac{U''(z)}{2U(z)}
  +\\ \frac{1}{4(a-z)^{2}(z-1)^{2}z^{2}}(4a^{2}k(1-k))+a^{2}\gamma(\gamma-2)+z^{4}(\gamma-3+\delta+\epsilon)(1+\gamma+\delta+\epsilon)+\\
  z(-8ak-4a^{2}k+8ak^{2}-2a^{2}(-2+\gamma(-2+\gamma+\delta))-2a\gamma(-2+\gamma+\epsilon))+\epsilon(2\gamma+\delta)+ \\
  z^{3}(4-4k-2(\gamma+\epsilon)(-2+\gamma+\delta+\epsilon)-2a(-3+\gamma^{2}+\delta(-2+\delta+\epsilon)+\gamma(-2+2\delta+\epsilon))))
  \textcolor[rgb]{1.00,0.00,0.00}{-}\textcolor[rgb]{1.00,0.00,0.00}{\frac{\alpha \beta z-q}{z(z-1)(z-a)}}=0.
  \end{split}
\end{equation}
After collecting the terms with respect to the powers of  $z^{i}, i=0,1,2,...$, one can find the parameters as
\begin{eqnarray}\label{z444}
  q&=& 2(1-k-2ka+2k^{2}a) \\ \label{z445}
  \delta &=& 1-2k,~~\epsilon=2 \\  \label{z446}
  \gamma\ &=& -2(k-1),~~\alpha=\frac{3+4k(k-2)}{\beta},~\beta=1.  \label{z447}
\end{eqnarray}
 So, (\ref{z40}) turns into
\begin{equation}\label{z44}
  Y''(z)+\left(\frac{1-2k}{z-1}-\frac{2(1-k)}{z}+\frac{2}{z-a}\right)Y'(z)+(\frac{(3+4k(k-2)z)-2+2k+4ka-4k^{2}a}{z(z-1)(z-1+k)})Y(z)=0.
\end{equation}
Again,  reminding Heun's equation (\ref{z41}) where $z=0, 1, a, \infty$ are the regular singular points satisfying the Fuchsian condition $\epsilon=\alpha+\beta-\gamma-\delta+1$. Matching (\ref{z41})  and (\ref{z44})  gives us (\ref{z444})-(\ref{z447}). On the other hand, our system (\ref{z44}) is two -parameters equation. In \cite{kara},
Nikiforov-Uvarov method is extended and applied to the general Heun  differential equation. According to the classification of the Heun equation given
in \cite{kara}, our system in (\ref{z44}) belongs to class-$V$ and using the results of \cite{kara}, we get
\begin{equation}\label{z46}
  Y_{1}(z)=z^{0}(z-1)^{0}(z-a)^{1-\epsilon}p(_{z}),
\end{equation}
where $p(z)$ is a polynomial which can be found as
\begin{equation}\label{pz}
  p(z)=a_1 (z-2k)z^{2k-1}+a_2 (z-1)^{2k},
\end{equation}
where $a_1, a_2$ are constants. Using Fuschian condition, we obtain two roots for the value of $k$:
\begin{equation}\label{kk}
  k=0,~~k=1.
\end{equation}
\\
\emph{\textbf{Case 3: Solutions in terms of Laguerre functions} }\\
If we apply the following transformation in (\ref{z11}) as
\begin{eqnarray}
  \bar{\phi}_{1}(r) &=& \frac{e^{-r/2}r^{3/2}}{r-M}Y(r) \\
 U(r) &=& C_2=const
\end{eqnarray}
where   $\lambda(r)=(1-\frac{M}{r})^{2}$, we get
\begin{equation}\label{z50}
\begin{split}
  rY''(r)+(1-r)Y'(r)+\left (\frac{1}{2}(-1+4C^{2}_{2}m-8C_2mM)-\frac{3}{4r}+\frac{1}{4}(1+4C^{2}_{2}-8C_2m)r\right)Y(r)=0.
\end{split}
\end{equation}
And $Y(r)$ can be found as
\begin{equation}\label{Y}
  Y(r) \sim C \exp\left[\left(\frac{1}{2}-i\sqrt{C_2}\sqrt{C_2-2m}\right)r\right]r^{\frac{\sqrt{3}}{2}} L_{n}\left(2i\sqrt{C_2}\sqrt{C_2-2m}r\right)
\end{equation}
where
\begin{equation}\label{s}
  n=-1-\sqrt{3}-2i\sqrt{C_2}\sqrt{C_2-2m},
\end{equation}
where $L_a(br)$ are the Laguerre polynomials. Then, $\bar{\phi}_1(r)$ becomes
\begin{equation}\label{fi}
  \bar{\phi}_{1}(r)=C \frac{1}{r-M}\exp\left[\left( -i\sqrt{C_2}\sqrt{C_2-2m}\right)r\right]r^{\frac{\sqrt{3}+3}{2}} L_{n}\left(2i\sqrt{C_2}\sqrt{C_2-2m}r\right)
\end{equation}
where $C$ is the normalization constant. To obtain the  real values of $n$, we can arrange   $C_2$ as :
\begin{equation}\label{C2}
  \sqrt{C_2}\sqrt{C_2-2m}=-iC_{3},~ \Rightarrow ~~ C_2=\frac{1}{2}(2m\pm \sqrt{C^{2}_{3}+4m^{2}}).
\end{equation}
By the way, the power of the exponential function should  vanish at infinity, then, one can take $C_2$ as
\begin{equation}\label{exp}
  C_2=\frac{1}{2}\left(2m + \sqrt{C^{2}_{3}+4m^{2}}\right).
\end{equation}
One can find $\varphi(r)$ vector potential component as
\begin{equation}\label{ff}
  \varphi(r)=\frac{1}{e}\left(m(1-\frac{M}{r})-E\right).
\end{equation}

\section{Conclusions}
The real eigenvalues and the corresponding spinor solutions are obtained for the Dirac equation for Reissner-Nordstr\"{o}m Black Holes. In our study, we take more general vector potential in the non-extremal and extremal black hole cases while the Coulombic $\sim \frac{1}{r}$ potentials were studied in the previous works \cite{9}. Our study may bring a new point of view to the bound state solutions for the partcile dynamics near  Reissner-Nordstr\"{o}m Black Holes. In this paper, we have also observed that the energy dependent potentials \cite{edp}, \cite{Z} which may appear in the bound state solutions. We have obtained the bound state solutions in terms of Airy, Kummer confluent hypergeometric and Whittaker functions in the case of non-extremal black holes while we have obtained those in terms of Bessel, Heun and Laguerre functions in the extremal case.

\end{document}